\documentclass[aps,prb,twocolumn,showpacs,groupedaddress]{revtex4}
\usepackage{graphicx}
\usepackage{bm}
\usepackage{bbm}
\usepackage{latexsym}
\usepackage{amssymb}
\usepackage{amsmath}

\begin{document}

\title{Thermodynamics of layered Heisenberg magnets with arbitrary spin}
\author{I. Juh\'asz Junger}
\author{D. Ihle}
\affiliation{Institut f\"{u}r Theoretische Physik,
Universit\"{a}t Leipzig, D-04109 Leipzig, Germany}
\author{J. Richter}
\affiliation{Institut f\"{u}r Theoretische Physik,
Otto-von-Guernicke-Universit\"{a}t Magdeburg, D-39016 Magdeburg, Germany}

\date{\today}

\begin{abstract}
We present a spin-rotation-invariant Green-function theory of long- and 
short-range order in the ferro- and antiferromagnetic Heisenberg model with 
arbitrary spin quantum number $S$ on a stacked square lattice. 
The thermodynamic 
quantities (Curie temperature $T_C$, N\'eel temperature $T_N$, specific heat 
$C_V$, intralayer and interlayer correlation lengths) are calculated, where the 
effects of the interlayer coupling and the $S$ dependence are explored. In 
addition, exact diagonalizations on finite two-dimensional (2D) lattices with 
$S \geqslant 1$ are performed, and a very good agreement between the results of 
both approaches is found. For the quasi-2D and isotropic 3D magnets, our 
theory agrees well with available quantum Monte Carlo and high-temperature 
series-expansion data. Comparing the quasi-2D $S=1/2$ magnets, we obtain the 
inequalities $T_N>T_C$ and, for small enough interlayer couplings, $T_N<T_C$. 
The results for $C_V$ and the intralayer correlation length are compared to 
experiments on the quasi-2D antiferromagnets Zn$_2$VO(PO$_4$)$_2$ with 
$S=1/2$ and La$_2$NiO$_4$ with $S=1$, respectively.                                             
\end{abstract} 
\maketitle

\section{INTRODUCTION}\label{sec:intro}
Low-dimensional ferromagnetic (FM) and antiferromagnetic (AF) quantum 
spin systems,\cite{SRF04} such as the quasi-two-dimensional (2D) Heisenberg 
ferromagnets [e.g., K$_2$CuF$_4$ with spin $S=1/2$ (Ref.~\onlinecite{LPS87})] 
and antiferromagnets [e.g., La$_2$NiO$_4$ with spin $S=1$ 
(Ref.~\onlinecite{NYH95}) being isostructural to the high-$T_C$ parent 
compound La$_2$CuO$_4$], are of current interest. Their study is motivated 
by the progress in the synthesis of new low-dimensional materials. For example, 
very recently a defective graphene sheet was reported to be a room-temperature 
ferromagnetic semiconductor that may be described by an effective 
quasi-2D Heisenberg model.\cite{PMH08} 

Investigations of layered Heisenberg magnets by numerical methods, e.g., 
quantum Monte Carlo (QMC) simulations and high-temperature series expansions 
(SE), have been performed for a selected number of cases and quantities only. 
QMC data are available for quasi-2D and spatially isotropic 3D antiferromagnets 
with $S=1/2$ (Refs.~\onlinecite{SSS03,YTH05,San98}) 
and $S=1$ (Ref.~\onlinecite{YTH05}). SE results exist for 
the 3D antiferromagnet with $S=1/2$, 1, and $3/2$ (Ref.~\onlinecite{OZ04}) 
and for the 3D ferromagnet with $S=1/2$ (Refs.~\onlinecite{OZ04} and 
\onlinecite{OB96}) and $S=1$ and $3/2$ (Ref.~\onlinecite{OZ04}). Note that 
numerical studies of ferromagnets and of $S>1/2$ systems are rather scarce. 

On the other hand, analytical approaches which are capable to evaluate 
the thermodynamics of layered ferro- and antiferromagnets with arbitrary 
spin below and above the magnetic transition temperature $T_M$ 
[$M=C,N$; $T_C$ $(T_N)$ denotes the Curie (N\'eel) temperature in the 
FM (AF) case] are desirable. In particular, the relation between $T_M$ 
and the relevant exchange couplings can be used to determine those couplings 
from experiments. Moreover, analytical theories may have the 
advantage of being applicable in such cases, where the QMC method cannot be 
applied, e.g., in the presence of frustration. However, the mean-field 
spin-wave theories based on the random-phase approximation 
(RPA),\cite{MSS92,DW94} that is equivalent to the Tyablikov decoupling of 
Green functions,\cite{Tya67} and on auxiliary-field representations 
(Schwinger-boson,\cite{IKK91, AA08} Dyson-Maleev,\cite{SI93} and 
boson-fermion representations\cite{IKK99}) are valid only at sufficiently 
low temperatures and do not adequately take into account the temperature 
dependence of magnetic short-range order (SRO) in the paramagnetic phase. 
For the 3D antiferromagnet, this deficiency has been removed by the quantum 
hierarchical reference theory of Ref.~\onlinecite{GP01}. For quasi-2D ferro- 
and antiferromagnets, an essential improvement in comparison to the standard 
mean-field approaches may be achieved by employing the second-order 
Green-function technique\cite{KY72} that we call, in the absence of spin 
anisotropies, rotation-invariant Green-function method (RGM). This 
technique provides a good description of SRO and long-range order (LRO) 
and has been applied recently successfully to low-dimensional quantum spin 
systems.\cite{SSI94,ISW99,YF00,SIH00,JIR04,APP07,JIB08,SRI04,JIR05,SRI05,
HRI08,MKB09} \vspace{-0.5mm}

In this paper we use the RGM and develop a theory of magnetic order 
in ferro- and antiferromagnets on a stacked square lattice. Thereby, we 
extend the previous work on the quasi-2D $S=1/2$ antiferromagnet\cite{SIH00} 
and the layered $S=1/2$ ferromagnet\cite{SRI05} to arbitrary values of the 
spin quantum number. 
We perform a systematic study of thermodynamic properties, where we 
contrast the FM with the AF cases. This allows to explore the role of 
quantum fluctuations.

We consider the 3D spatially anisotropic Heisenberg model with arbitrary 
spin $S$,
\begin{equation}
H=\frac{J_{\parallel}}{2} \sum_{\langle i,j \rangle_{xy}}  \bm{S}_i \bm{S}_j
  +\frac{J_{\perp}}{2}  \sum_{\langle i,j \rangle_{z}}   \bm{S}_i \bm{S}_j
\label{eq_ham}
\end{equation}
[$\langle i,j \rangle_{xy}$ and $\langle i,j \rangle_{z}$ denote 
nearest-neighbor (NN) sites in the $xy$ plane and along the $z$ direction 
of a simple cubic lattice, respectively] with $\bm{S}_i^2=S(S+1)$. For the 
layered ferromagnet (antiferromagnet) we have $J_{\mu} < 0$ ($J_{\mu} > 0$), 
where $\mu= \parallel$, $\perp$. We calculate the thermodynamic properties 
(magnetic transition temperatures, specific heat, and correlation lengths) and 
study the crossover from isotropic 2D ($J_{\perp}=0$) to 3D 
($J_{\perp}=J_{\parallel}$) quantum magnets. For comparison, we perform 
Lanczos exact diagonalizations (ED) to calculate the ground state  
of the 2D antiferromagnet with $S=1$, $\frac{3}{2}$, and 2 
on a lattice of $N=16$ sites and full ED to get the thermodynamic 
quantities for the 2D $S=1$ ferromagnet on a lattice of 
$N=8$ sites. 

The rest of the paper is organized as follows: In Sec.~\ref{sec:gfmet}, 
the theory based on the RGM for model (\ref{eq_ham}) is developed, 
where the extension of 
previous RGM approaches \cite{SIH00, SRI05} to arbitrary spins implies 
novel technical aspects. In Sec.~\ref{sec:res}, the thermodynamic properties 
of the 2D and 3D ferromagnets and antiferromagnets are investigated as 
functions of temperature, spin, and interlayer coupling, also in comparison 
to available QMC and SE data, 
and are related to experiments. Finally, a summary of our work is given 
in Sec.~\ref{sec:sum}.
\section{ROTATION-INVARIANT GREEN-FUNCTION THEORY}\label{sec:gfmet}
To evaluate the spin-correlation functions and the thermodynamic quantities,                                                 
we calculate the dynamic spin susceptibility 
$\chi_{\bm{q}}^{+-} (\omega)= -
\langle \langle S_{\bm{q}}^{+}; S_{-\bm{q}}^{-} \rangle \rangle_{\omega} $ 
 (here, $\langle \langle \ldots ; \ldots \rangle \rangle_{\omega}$ 
 denotes the two-time 
commutator Green function\cite{Tya67}) by the RGM.\cite{KY72} Using the 
equations of motion up to the second step and supposing rotational symmetry 
in spin space, i.e., $\langle S_i^z \rangle=0 $,  we obtain 
$ \omega^2 
\langle \langle S_{\bm{q}}^{+}; S_{-\bm{q}}^{-} \rangle \rangle_{\omega} = 
M_{\bm{q}} +
\langle \langle -\ddot{S}_{\bm{q}}^{+}; S_{-\bm{q}}^{-} 
\rangle \rangle_{\omega}$  
with
$ M_{\bm{q}}= \left\langle \left[ [S_{\bm{q}}^{+},H], S_{-\bm{q}}^{-}\right] 
\right \rangle $ 
and 
$ -\ddot{S}_{\bm{q}}^{+}= \left[ [S_{\bm{q}}^{+},H], H \right] $. 
For the model (\ref{eq_ham}) the moment $M_{\bm{q}}$ is given by the 
exact expression  
\begin{equation}
M_{\bm{q}}= - 8 J_{\parallel} C_{100} (1-\gamma_{\bm{q}})
            - 4 J_{\perp}   C_{001} (1-\cos q_z),
\label{eq_mq}
\end{equation}
where
$C_{mnl} \equiv C_{\bm{R}}=\langle S_{0}^{+} S_{\bm{R}}^{-} \rangle
= 2 \langle S_{0}^{z} S_{\bm{R}}^{z} \rangle$,  
$\bm{R}=m \bm{e}_x + n \bm{e}_y + l \bm{e}_z$, 
and 
$\gamma_{\bm{q}}=\frac{1}{2} (\cos q_x + \cos q_y)$. 
The second derivative $-\ddot{S}_{\bm{q}}^{+}$ is approximated in the spirit of 
the schemes employed in Refs.~\onlinecite{KY72, ISW99, SIH00, JIB08}, 
and \onlinecite{SRI04}. 
That means, in $-\ddot{S}_{i}^{+}$ we decouple the  
products of three spin operators along NN sequences 
$\langle i, j, l \rangle$
as
\begin{equation}
S_i^+ S_j^+ S_l^-= \alpha_{ 1 \mu} \langle S_j^+ S_l^- \rangle S_i^+ + 
		   \alpha_{ 2 \mu} \langle S_i^+ S_l^- \rangle S_j^+ ,	
\label{eq_entk1}
\end{equation}
where the vertex parameters $\alpha_{ 1 \mu}$ and $\alpha_{ 2 \mu}$ 
are attached to NN and further-distant correlation functions, 
respectively, either within a layer ($\mu = \parallel$) or between two 
layers ($\mu = \perp $). The products of three spin operators with two 
coinciding sites, appearing for $S \geqslant 1$, are decoupled 
as\cite{SSI94,JIR05,JIB08}
\begin{equation}
S_i^+ S_j^- S_j^+= \langle S_j^- S_j^+ \rangle S_i^+ + 
		   \lambda_{\mu} \langle S_i^+ S_j^- \rangle S_j^+ ,	
\label{eq_entk2}
\end{equation}
where the vertex parameter $\lambda_{\mu}$ is associated with the NN 
correlator in the layer or between NN layers. We obtain
$-\ddot{S}_{\bm{q}}^+=\omega_{\bm{q}}^2 S_{\bm{q}}^+$
and
\begin{equation}
\chi_{\bm{q}}^{+-} (\omega)=
-\langle \langle S_{\bm{q}}^{+} ; S_{-\bm{q}}^{-} \rangle \rangle_{\omega}=
\frac{M_{\bm{q}}}{\omega_{\bm{q}}^2-\omega^2},
\label{eq_gf}
\end{equation}
with
\begin{eqnarray}
\omega_{\bm{q}}^2&=&(1-\gamma_{\bm{q}}) 
\{ \Delta_{\parallel}
+ 16 J_{\parallel}^2 \alpha_{1 \parallel} C_{100} (1-\gamma_{{\bm{q}}}) \} 
\nonumber \\
&& +(1-\cos q_z) 
\{ \Delta_{\perp}+ 4 J_{\perp}^2 \alpha_{1 \perp} C_{001} (1-\cos q_z) \} 
\label{eq_omega_q}
\nonumber \\
&& + \tilde{\Delta} (1-\gamma_{{\bm{q}}}) (1-\cos q_z), 
% \label{eq_omega_q}
\end{eqnarray}
%
% Delta_xy
\begin{eqnarray}
\Delta_{\parallel}&=&2 J_{\parallel}^2 
\{\bar{S} +2 \lambda_{\parallel} C_{100} 
+ 2 \alpha_{2 \parallel} (2 C_{110} + C_{200})  \\
&-& 10 \alpha_{1 \parallel} C_{100} \} 
% \nonumber \\
+ 8 J_{\parallel} J_{\perp} (\alpha_{2 \perp} C_{101}
-\alpha_{1 \parallel} C_{100}) , \nonumber
\label{eq_delta_xy}
\end{eqnarray}
%
% Delta_z
\begin{eqnarray}
\Delta_{\perp}&=& J_{\perp}^2 
\{\bar{S} +2 \lambda_{\perp} C_{001} 
+ 2 \alpha_{2 \perp} C_{002} -6 \alpha_{1 \perp} C_{001} \} 
\nonumber \\
&+& 8 J_{\parallel} J_{\perp} 
(\alpha_{2 \perp} C_{101}-\alpha_{1 \perp} C_{001}) ,
\label{eq_delta_z}
\end{eqnarray}
%
% Delta_xyz
\begin{equation}
\tilde{\Delta} = 8 J_{\parallel} J_{\perp} (\alpha_{1 \parallel} C_{100}
             + \alpha_{1 \perp} C_{001}), 
\label{eq_delta_xyz}
\end{equation}
where $\bar{S}=\frac{4}{3} S(S+1)$. 
From the Green function (\ref{eq_gf}) the correlation functions  
$C_{\bm{R}}=\frac{1}{N} \sum_{\bm{q}} C_{\bm{q}} 
\text{e}^{i\bm{qR}} 
$
are determined by the spectral theorem,\cite{Tya67}
\begin{equation}
C_{\bm{q}}=\langle S_{\bm{q}}^+  S_{-\bm{q}}^- \rangle 
=\frac{M_{\bm{q}}}{2 \omega_{\bm{q}}} [1+2 n(\omega_{\bm{q}})] , 
\label{eq_C_q}
\end{equation} 
where 
$ n(\omega)=(\text{e}^{\omega/T}-1)^{-1} $
is the Bose function. The NN correlators are directly related to the internal 
energy $u$ per site, 
$u=3 J_{\parallel} C_{100} + \frac{3}{2} J_{\perp} C_{001} $,
from which the specific heat
$C_{V}=du/dT$
may be calculated. Taking the on-site correlator $C_{\bm{R}=0}$ and using the 
operator identity 
$\bm{S}_{i}^{2}=S_i^+ S_i^- - S_i^z + (S_i^z)^2$,
we get the sum rule  
\begin{equation}
\frac{1}{N} \sum_{\bm{q}} C_{\bm{q}}=\frac{2}{3} S(S+1).
\label{eq_sr}
\end{equation}
Let us consider the static spin susceptibility 
$\chi_{\bm{q}} \equiv \chi_{\bm{q}} (\omega=0) $
with
$\chi_{\bm{q}} (\omega) \equiv \chi_{\bm{q}}^{zz} (\omega) =  
\frac{1}{2} \chi_{\bm{q}}^{+-} (\omega)$,
i.e.,
$\chi_{\bm{q}} = M_{\bm{q}} / 2 \omega_{\bm{q}}^{2}$. The lowest-order 
expansion of $M_{\bm{q}}$ and $ \omega_{\bm{q}}^{2}$ at $\bm{q}=0$ yields 
$
\chi_{\bm{q}} =[a (q_x^2+q_y^2)+b q_z^2]/[c (q_x^2+q_y^2)+d q_z^2], 
$
where
$a=- J_{\parallel} C_{100}$,  $ b=- J_{\perp} C_{001}$,  
$c=\Delta_{\parallel}/4$,
and $d=\Delta_{\perp}/2$.   
Calculating the uniform static susceptibility 
$\chi = \lim_{\bm{q} \to 0} \chi_{\bm{q}}$, the ratio of the anisotropic 
functions $M_{\bm{q}}$ and $\omega_{\bm{q}}^{2}$ must be isotropic in the 
limit $\bm{q} \to 0$, i.e., 
$\lim_{q_{x(y)} \to 0} \chi_{\bm{q}} \vert_{q_z=0}=
\lim_{q_{z} \to 0} \chi_{\bm{q}} \vert_{q_{x(y)}=0}.$
That is, the condition $a/c=b/d$ has to be fulfilled which reads 
as the isotropy condition
\begin{equation}
\chi=
-\frac{4 }{\Delta_{\parallel}} J_{\parallel} C_{100} =
-\frac{2 }{\Delta_{\perp}} J_{\perp} C_{001}.
\label{eq_iso}
\end{equation}
Note that such a condition was also employed in 
Refs.~\onlinecite{ISW99}, \onlinecite{SIH00}, \onlinecite{SRI04}, 
and \onlinecite{SRI05}.

The phase with magnetic LRO at $T \leqslant T_M$ 
is described by the divergence of the static susceptibility 
at the ordering vector $\bm{q}_0$, i.e., by $\chi_{\bm{q}_0}^{-1}=0$, 
with $\bm{q}_0=0$ and $\bm{q}_0 = \bm{Q}=(\pi,\pi,\pi)$ in the FM and AF 
case, respectively. In this phase the correlation function $C_{\bm{R}}$ 
is written as\cite{KY72}
\begin{equation}
C_{\bm{R}}=\frac{1}{N} \sum_{\bm{q} (\neq \bm{q}_0 )} C_{\bm{q}} 
\text{e}^{i\bm{qR}} 
+ C \text{e}^{i\bm{q}_0 \bm{R}}
\label{eq_C_R}
\end{equation} 
with $C_{\bm{q}}$ given by Eq.~(\ref{eq_C_q}). The condensation part $C$ 
determines the magnetization $m$ that is defined in the 
spin-rotation-invariant form 
$
m^2=\frac{3}{2N} \sum_{\bm{R}} C_{\bm{R}} \text{e}^{-i \bm{q}_0 \bm{R}}
   =\frac{3}{2} C 
$. 
The LRO conditions for the ferromagnet and antiferromagnet 
read as $\Delta_{\mu}=0$ [cf. Eq.~(\ref{eq_iso})] and 
$\omega_{\bm{Q}}=0$, respectively. 

The magnetic correlation lengths above $T_M$ may be calculated by 
expanding $\chi_{\bm{q}}$ in the neighborhood of the vector 
$\bm{q}_0$.\cite{KY72,JIB08,HRI08} For the ferromagnet ($\bm{q}_0=0$),  
the expansion yields 
$\chi_{\bm{q}}= \chi [1+\xi_{\parallel}^{2} (q_x^2+q_y^2) + 
\xi_{\perp}^2 q_z^2]^{-1}$ with the squared intralayer ($\mu=\parallel$) 
and interlayer ($\mu = \perp$) correlation lengths
\begin{equation}
\xi_{\mu}^2=|J_{\mu} | \alpha_{1 \mu} \chi.
\label{eq_fm_ksi} 
\end{equation}
For the antiferromagnet, the expansion around $\bm{q}_0=\bm{Q}$ gives 
$\chi_{\bm{q}}= \chi_{\bm{Q}} [1+\xi_{\parallel}^{2} (k_x^2+k_y^2) + 
\xi_{\perp}^2 k_z^2]^{-1}$
with $\bm{k}=\bm{q}-\bm{Q}$ and 
\begin{equation}
\xi_{\parallel}^{2}=
-\frac{1}{4 \omega_{\bm{Q}}^{2}} 
(\Delta_{\parallel} +64 J_{\parallel}^2 \alpha_{1 \parallel} C_{100} 
+ 2 \tilde{\Delta})
-\frac{2 J_{\parallel} C_{100}}{M_{\bf{Q}}}, 
\label{eq_afm_ksi_xy}
\end{equation}
\begin{equation}
\xi_{\perp}^{2}=
- \frac{1}{2 \omega_{\bm{Q}}^2}
(\Delta_{\perp} +16 J_{\perp}^2 \alpha_{1 \perp} C_{001}+ 2 \tilde{\Delta})
-\frac{2 J_{\perp} C_{001}}{M_{\bf{Q}}} .
\label{eq_afm_ksi_z}
\end{equation}

To evaluate the thermodynamic properties, the correlation functions 
$C_{\bm{R}}$ and the vertex parameters $\alpha_{1 \mu} $, 
$\alpha_{2 \mu} $, and $\lambda_{\mu}$ appearing in the spectrum 
$\omega_{\bm{q}}$ [Eqs.~(\ref{eq_omega_q})-(\ref{eq_delta_xyz})] 
as well as the condensation term $C$ in the LRO phase have to be 
determined. Besides Eqs.~(\ref{eq_C_q}) and (\ref{eq_C_R}) for calculating 
the correlators, we have the sum rule (\ref{eq_sr}), 
the isotropy condition (\ref{eq_iso}), and the LRO conditions for 
determining the parameters; that is, we have more parameters than 
equations. To obtain a closed system of self-consistency equations, 
we reduce the number of parameters by reasonable simplifications 
that we have to specify for the FM and AF cases.
\\
(i) \textit{Ferromagnet}: Considering the ground state ($T=0$), we 
have the exact result
\begin{equation}
C_{\bm{R}} (0)=\frac{2}{3}S \delta_{\bm{R},0} + \frac{2}{3} S^2,
\label{eq_ex} 
\end{equation}
which can be reproduced by Eq.~(\ref{eq_C_R}), 
$C_{\bm{R}} (0)= \frac{1}{N} \sum_{\bm{q}(\neq 0)} 
[M_{\bm{q}}(0)/2 \omega_{\bm{q}}(0)] \text{e}^{i \bm{qR}} + C(0)$, 
if $C(0)=\frac{2}{3} S^2$ 
and $M_{\bm{q}}(0)/2 \omega_{\bm{q}}(0)=\frac{2}{3} S$. 
The equality  
$M_{\bm{q}}^2(0)=\frac{16}{9}S^2 \omega_{\bm{q}}^2 (0)$ requires the 
equations $\alpha_{1 \mu}(0)=\frac{3}{2}$ and $\Delta_{\mu}(0)=0$ 
(LRO condition, see above) or, explicitly,
$
J_{\parallel} \left(  1+ \frac{1}{S} + \lambda_{\parallel} 
+ 3 \alpha_{2 \parallel} -\frac{15}{2} \right)  
+ 2 J_{\perp} \left( \alpha_{2 \perp}-\frac{3}{2}\right) = 0 
$ 
and
$
J_{\perp} \left(  1+ \frac{1}{S} + \lambda_{\perp} 
+ \alpha_{2 \perp} -\frac{9}{2} \right)  
+ 4 J_{\parallel} \left( \alpha_{2 \perp}-\frac{3}{2}\right) =0
$.
In the special case $S=1/2$, in $-\ddot{S}_{i}^{+}$, products of spin 
operators with two coinciding sites do not appear, which is equivalent 
to setting $\lambda_{\mu}=0$. Then, the solution of the equations 
$\Delta_{\mu} (0)=0$ yields $\alpha_{2 \mu} (0) = \frac{3}{2}$, i.e., we have 
$\alpha_{2 \mu}(0)=\alpha_{1 \mu}(0)$. We take this equality also for 
$S \geqslant 1$ and get $\lambda_{\mu}(0)=2-\frac{1}{S}$. To determine 
the parameters at finite temperatures, we first consider the high-temperature 
limit, where all $\alpha$ parameters approach unity,\cite{KY72} 
$\lim_{T \to \infty} \alpha_{1,2 \mu} (T)=1$, 
and the high-temperature series expansion\cite{JIR05} yields 
$\lim_{T \to \infty} \lambda_{\mu}(T) \equiv \lambda_{\infty} =
1-3[4S(S+1)]^{-1}$. Because we have identical vertex parameters 
$\alpha_{2 \mu}$ and $\alpha_{1 \mu}$ as well as identical parameters 
$\lambda_{\parallel}$ 
and $\lambda_{\perp}$ at $T=0$ and for $T \to \infty$, we put 
$\alpha_{2 \mu} (T)=\alpha_{1 \mu} (T) \equiv \alpha_{\mu} (T)$ and   
$\lambda_{\parallel} (T)=\lambda_{\perp} (T) \equiv \lambda (T)$ in the 
whole temperature region. Then, at $T \leqslant T_C$ we have the four 
parameters $\alpha_{\parallel}$, $\alpha_{\perp}$, $\lambda$ and $C$. 
For their determination, besides the sum rule (\ref{eq_sr}) and the LRO 
conditions, $\Delta_{\parallel}=0$ and $\Delta_{\perp}=0$, we need an 
additional condition. Reasoning similarly as in Ref.~\onlinecite{KY72} 
for $\alpha$ parameters, we consider the ratio 
\begin{equation}
r_{\lambda}(T) \equiv
\frac{\lambda (T)-\lambda_{\infty}}
{\alpha_{\parallel}(T)- 1} = r_{\lambda}(0)
\label{eq_r_lambda}
\end{equation}
as temperature independent. For $T > T_C$ $(C=0)$ we have $\Delta_{\mu}>0$, 
and the number of quantities and equations 
[Eqs.~(\ref{eq_sr}), (\ref{eq_iso}), (\ref{eq_r_lambda})] is reduced by one. 
\\
(ii) \textit{Antiferromagnet:} As revealed by previous studies of the 
2D $S=1/2$ antiferromagnet,\cite{KY72} 
contrary to the FM case, the introduction of the vertex 
parameter $\alpha_{2} \neq \alpha_{1}$ appreciably improves the results 
as compared with the simplification 
$\alpha_{2} = \alpha_{1}$. 
We expect the same behavior also for the layered antiferromagnet. 
This can be understood as follows. In the LRO phase and paraphase with AF SRO, 
the parameter $\alpha_{1 \mu}$ is associated with NN correlators of negative 
sign, whereas $\alpha_{2 \mu}$ is connected with positive further-distant 
correlation functions. Therefore, the difference in the sign of the correlators 
may be the reason for the relevance of the difference between 
$\alpha_{1 \mu}$ 
and $\alpha_{2 \mu}$. This is in contrast to the FM case, where all 
correlators have a positive sign, and the equality 
$\alpha_{2 \mu}=\alpha_{1 \mu}$ is a good assumption. 
Accordingly, we put 
$\alpha_{2 \mu} = \alpha_{2}$ (cf. Ref.~\onlinecite{SIH00}), and, as in the 
FM case, we take $\lambda_{\mu}=\lambda$. To determine the five parameters 
$\alpha_{1 \parallel}$, $\alpha_{1 \perp}$, $\alpha_{2}$, $\lambda$ and 
$C$ at $T=0$, we have the sum rule (\ref{eq_sr}), the isotropy condition 
(\ref{eq_iso}), and the LRO condition $\omega_{\bm{Q}}=0$. As the two 
additional conditions for fixing the free parameters, we assume 
$\lambda (0)$ to be equal to the FM value, i.e., $\lambda(0)=2-\frac{1}{S}$, 
and adjust the ground-state energy $u(0)$ to the expression given by the 
linear spin-wave theory (LSWT), 
$u(0)=u_{LSWT}(0)= 
-S(S+1) (2 J_{\parallel}+J_{\perp})+
\frac{S}{N} \sum_{\bm{q}} \sqrt{(2 J_{\parallel}+J_{\perp})^2
-(2 J_{\parallel} \gamma_{\bm{q}} + J_{\perp} \cos q_z)^2}
$. 
At finite temperatures, besides Eqs.~(\ref{eq_sr}) and (\ref{eq_iso}), and 
$\omega_{\bm{Q}}=0$ (for $T \leqslant T_{N}$), we take 
Eq.~(\ref{eq_r_lambda}) with $\alpha_{\parallel}(T)$ replaced by 
$\alpha_{1 \parallel}(T)$ and the analogous condition 
(cf. Refs.~\onlinecite{KY72} and \onlinecite{SIH00}) 
\begin{equation}
r_{\alpha}(T)\equiv
\frac{\alpha_2(T)-1}{\alpha_{1\parallel}(T)-1}=r_{\alpha}(0).
\label{eq_r_alpha} 
\end{equation}

\section{RESULTS}\label{sec:res} 
As described in Sec.~\ref{sec:gfmet}, the quantities of the RGM determining the 
thermodynamic properties have to be numerically calculated as solutions of a 
coupled system of nonlinear algebraic self-consistency equations. For example, 
considering the 
antiferromagnet at $T \leqslant T_N$, we have 11 equations for $C_{100}$, 
$C_{001}$, $C_{110}$, $C_{200}$, $C_{101}$, $C_{002}$ 
[appearing in Eqs.~(\ref{eq_omega_q})-(\ref{eq_delta_xyz}) and calculated 
by Eq.~(\ref{eq_C_R})], 
$\alpha_{1 \parallel}$, $\alpha_{1 \perp}$, $\alpha_{2}$, $\lambda$ and 
$C$. To solve this system of equations, we use Broyden's method,\cite{PTV01} 
which yields the solutions with a relative error of about $10^{-7}$ on the 
average. The momentum integrals occurring in the self-consistency equations 
are done by Gaussian integration.
\subsection{Two-dimensional $S \geqslant 1$ magnets}

 \begin{table}[b]
 \begin{minipage}{18cm}
 \caption{Correlation functions $C_{\bm{R}}$ of the 2D antiferromagnet 
 at $T=0$, as obtained by the RGM in the thermodynamic limit and for a 
 finite system with $N=16$, denoted by RGM(16), in comparison with the ED 
 data for $N=16$. }
 \begin{center}
 \begin{tabular}{|c|ccc|ccc|ccc|} \hline \hline
 & \multicolumn{3}{c|}{S=1} & \multicolumn{3}{c|}{S=3/2} 
 & \multicolumn{3}{c|}{S=2} \\
 \phantom{n}$\vec{R}$\phantom{nn}
 & \phantom{n}RGM\phantom{n} & \phantom{n}RGM(16)\phantom{n}
 & \phantom{n}ED\phantom{n}
 & \phantom{n}RGM\phantom{n} & \phantom{n}RGM(16)\phantom{n}
 & \phantom{n}ED\phantom{n}
 & \phantom{n}RGM\phantom{n} & \phantom{n}RGM(16)\phantom{n}
 & \phantom{n}ED\phantom{n} \\ \hline
 (1,0)& -0.7720 & -0.7947 & -0.7980 &-1.6579 & -1.6920 & -1.6954 
 & -2.8773 & -2.9227 & -2.9261  \\
 (1,1)&  0.5985 &  0.6156 &  0.6169 & 1.3977 &  1.4230 &  1.4242 
 &  2.5303 &  2.5638 &  2.5650  \\
 (2,1)& -0.5406 & -0.6032 & -0.6029 &-1.3109 & -1.4040 & -1.4035 
 & -2.4146 & -2.5383 & -2.5376  \\
 (2,2)&  0.5077 &  0.5649 &  0.5689 & 1.2616 &  1.3462 &  1.3503 
 &  2.3488 &  2.4611 &  2.4651  \\ \hline \hline
 \end{tabular}
 \end{center}
 \label{tab_1}
 \end{minipage}
\end{table}

To test the quality of the approximations made in the RGM, in particular 
the assumptions about the vertex parameters introduced in the decouplings 
(\ref{eq_entk1}) and (\ref{eq_entk2}), we consider some correlation functions 
and thermodynamic properties of 2D $S \geqslant 1$ magnets in comparison 
with ED and QMC data. To provide a better comparison of the RGM with ED 
results, we apply the RGM also to finite systems with periodic boundary 
conditions proceeding as in Ref.~\onlinecite{JIR05}. 
In Table \ref{tab_1} our RGM and 
ED results for several correlation functions of the 2D antiferromagnet at 
$T=0$, also obtained by the RGM for a $N=4 \times 4$ square lattice, 
are presented. Determining 
the parameters (see Sec.~\ref{sec:gfmet}) for the finite system with $N=16$, 
as an input we take the ground-state energy in the LSWT that is also evaluated 
for $N=16$.
Let us consider the NN correlator $C_{10}(0)$ determining the ground-state 
energy $u(0)=3 J_{\parallel} C_{10}$. The LSWT and ED results are in a good 
agreement (for $S=2$ they differ by only 0.1\%). This provides some 
justification for using the LSWT data for $u(0)$ as an input also in 
the 3D 
AF case. Note that the LSWT input is of advantage as compared with the 
choice made in Ref.~\onlinecite{SIH00}, where $u(0)$ is composed 
approximately from 1D and 2D energy contributions which is justified 
for $J_{\perp}/J_{\parallel} \ll 1$ only. The further-distant correlators 
listed in Table \ref{tab_1} and calculated by the RGM for $N=16$ agree 
remarkably well (with an average deviation of 0.2\%) with the ED results.

\begin{figure}[t]
   \centering
   \includegraphics[scale=0.65]{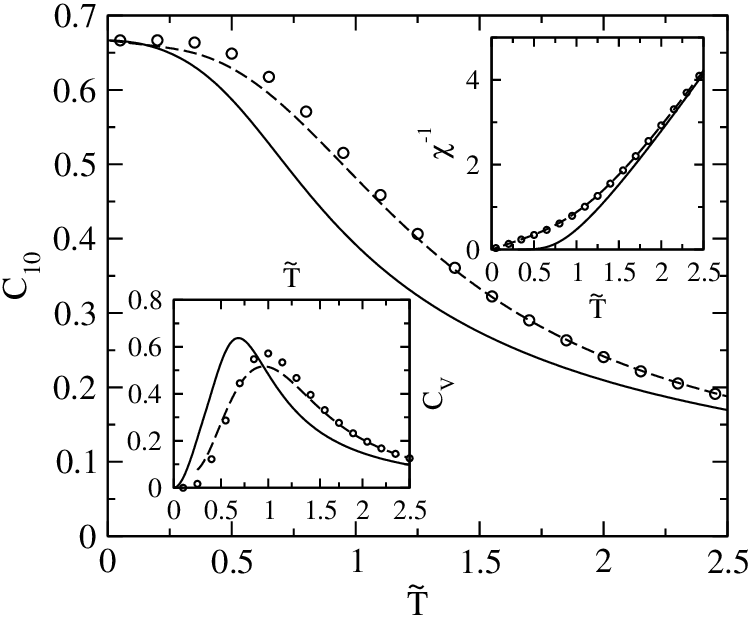}
   \caption{2D $S=1$ ferromagnet: NN correlation function 
   $C_{10}$, uniform static susceptibility $\chi$ (upper inset), and 
   specific heat $C_V$ (lower inset) as functions of 
   $\tilde{T}=T/[|J_{\parallel}| S(S+1)]$, where the results of the RGM 
   in the thermodynamic limit (solid lines) and for $N=8$ (dashed lines) 
   and the ED data ($\circ$, $N=8$) are shown.  }
   \label{fig_2d_fm}
\end{figure}

Considering the 2D $S=1$ ferromagnet, in Fig.~\ref{fig_2d_fm}   
the temperature dependence of $C_{10}$, $\chi$, and $C_V$  is 
plotted. For the finite lattice with $N=8$, a very good agreement 
of the RGM and ED data is found. 
The comparison with the RGM results for $N \to \infty$ demonstrates the 
finite-size effects.

\begin{figure}[t]
   \centering
   \includegraphics[scale=0.65]{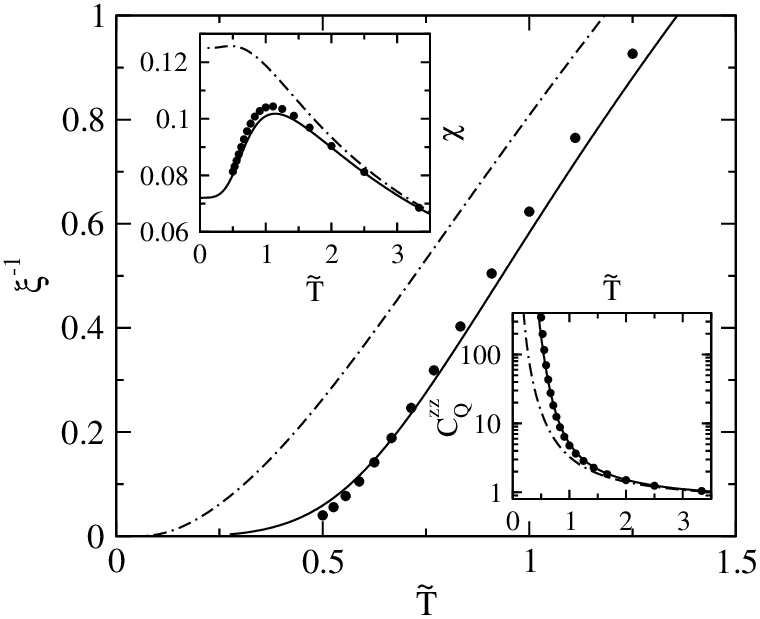}
   \caption{2D $S=1$ antiferromagnet: Correlation length 
   $\xi$, uniform static susceptibility $\chi$ (upper inset), and 
   staggered structure factor $C_{\bm{Q}}^{zz}=\frac{1}{2} C_{\bm{Q}}$ 
   (lower inset) as functions of 
   $\tilde{T}=T/[J_{\parallel} S(S+1)]$, where the RGM results (solid lines) 
   are compared with the QMC data of Ref.~\onlinecite{HTK98} ($\bullet$). 
   For comparison, the results of a simplified version of the RGM with 
   $\lambda(T)=\lambda(0)$ (see text) are depicted (dot-dashed lines).
   }
   \label{fig_2d_afm}
\end{figure}

Next, we consider the 2D antiferromagnet at finite temperatures. 
Since the case $S=1/2$ was 
intensively studied by the RGM in previous work,\cite{KY72,ISW99} we 
compare our results for $S=1$ with available QMC 
data.\cite{HTK98} As can be seen in Fig.~\ref{fig_2d_afm}, we obtain a 
surprisingly good agreement of the RGM with the QMC results (note that the QMC 
data for the correlation length agree with the SE results of 
Ref.~\onlinecite{ESS95}). This agreement is much better than for the 
$S=1/2$ antiferromagnet.\cite{KY72} Correspondingly, for $S=1$ we can give 
a rather reliable value for the zero-temperature susceptibility, 
$\chi(0)=0.07197$.

As outlined in Sec.~\ref{sec:gfmet}, in our approach more vertex 
parameters are introduced as independent equations for 
\newpage \noindent
them can be provided by the RGM. 
Therefore, we have to formulate appropriate additional 
conditions for their determination. Let us discuss, in comparison to the 
choice fixed in Sec.~\ref{sec:gfmet}, two alternate choices of the 
parameters 
$\alpha_2$ and $\lambda$ for the 2D $S=1$ antiferromagnet 
(in two dimensions we omit the index $\mu=\parallel$, e.g., 
$\alpha_{1,2 \parallel}=\alpha_{1,2}$), which are analogous to the choices 
made previously for the $S=1/2$ antiferromagnet,\cite{KY72} and the 
$S \geqslant 1$ ferromagnet.\cite{SSI94} 
(i) If we choose 
$\alpha_2=\alpha_1$, the parameter $\lambda(0)$ 
can be calculated (note that $\alpha_2$ and $\lambda$ only appear in the 
combination given by $\Delta$)  
 and used in Eq.~(\ref{eq_r_lambda}). Then, we find 
the finite-temperature results 
 to be not in such a good agreement with the QMC data as the results 
obtained by the parameter choice with $\alpha_2 \neq \alpha_1$. 
This corresponds to the findings for the 
$S=1/2$ antiferromagnet\cite{KY72} 
and may be understood as explained in Sec.~\ref{sec:gfmet}. Therefore, 
we discard the choice $\alpha_2=\alpha_1$.
(ii) If we adopt $\alpha_2 \neq \alpha_1$, 
but neglect the temperature dependence of $\lambda$, i.e., 
$\lambda(T)=\lambda(0)=2-\frac{1}{S}$ (as was assumed for the FM case 
in Ref.~\onlinecite{SSI94}), the results appreciably deviate from the 
QMC data, as is demonstrated in Fig.~\ref{fig_2d_afm} (dot-dashed lines). 
This gives strong arguments 
for taking into account the decrease 
of $\lambda(T)$ with increasing temperature [e.g., for $S=1$, we have 
$\lambda(0)=1$ and $\lambda_{\infty}=0.625$] and for 
our choice of the parameters for the 
antiferromagnet outlined on Sec.~\ref{sec:gfmet}. Note that for the 
$S=1$ ferromagnet, where $\alpha_2=\alpha_1$, the results shown in 
Fig.~\ref{fig_2d_fm} only slightly improve those obtained by the 
assumption $\lambda(T)=\lambda(0)$. 
\subsection{Transition temperatures}\label{subsec:Tcn}
\begin{table*}%[t]
\caption{Transition temperatures $\tilde{T}_M=T_M/[|J_{\parallel}|S(S+1)]$ 
of the ferromagnet ($\tilde{T}_C$) and antiferromagnet ($\tilde{T}_N$) 
calculated by the RGM for different spins $S$ and interlayer couplings   
$J_{\perp}/J_{\parallel}$.} 
\begin{center}
\begin{tabular}{c|ccc|ccc|c} \hline \hline
& \multicolumn{3}{c|}{Ferromagnet} & \multicolumn{3}{c|}{Antiferromagnet} \\
\phantom{n}$J_{\bot}/J_{||}$\phantom{nn}
& \phantom{n} $S=1/2$ \phantom{n} & \phantom{n} $S=1$ \phantom{n} 
& \phantom{n} $S=3/2$ \phantom{n} & \phantom{n} $S=1/2$ \phantom{n}
& \phantom{n} $S=1$ \phantom{n} & \phantom{n} $S=3/2$ \phantom{n} 
& \phantom{n} $S=\infty$ \phantom{n} \\
\hline
0.0001 & 0.2457 & 0.3243 & 0.3542 & 0.1589 & 0.3393 & 0.3681 & 0.3305 \\
0.0005 & 0.2928 & 0.3758 & 0.4041 & 0.2150 & 0.4014 & 0.4170 & 0.3785 \\
0.001  & 0.3184 & 0.4027 & 0.4298 & 0.2498 & 0.4331 & 0.4421 & 0.4039 \\
0.005  & 0.3961 & 0.4803 & 0.5035 & 0.3694 & 0.5195 & 0.5133 & 0.4784 \\
0.01   & 0.4403 & 0.5226 & 0.5436 & 0.4430 & 0.5640 & 0.5521 & 0.5200 \\
0.02   & 0.4935 & 0.5725 & 0.5911 & 0.5311 & 0.6150 & 0.5986 & 0.5698 \\
0.05   & 0.5826 & 0.6552 & 0.6706 & 0.6681 & 0.6979 & 0.6776 & 0.6538 \\
0.1    & 0.6699 & 0.7368 & 0.7503 & 0.7870 & 0.7800 & 0.7583 & 0.7378 \\
0.5    & 0.9953 & 1.0571 & 1.0694 & 1.1655 & 1.1121 & 1.0884 & 1.0667 \\
1.0    & 1.2346 & 1.3063 & 1.3208 & 1.4382 & 1.3762 & 1.3478 & 1.3189 \\
\hline \hline
\end{tabular}
\end{center}
\label{tab_2}
\end{table*}

An important problem in the study of layered ferromagnets and 
antiferromagnets is the calculation of the transition temperature 
$T_M$ ($M=C, N$) as a function of the interlayer coupling $J_{\perp}$ 
and of the spin quantum number $S$. From the experimental side, the 
knowledge of the dependence $T_M (R,S)$ with $R=J_{\perp}/J_{\parallel}$ 
is useful to estimate the interlayer exchange coupling from measurements  
of $T_M$. To test the quality of analytical approaches, the precise results 
of numerical methods, such as the QMC\cite{YTH05} and SE data,\cite{OZ04} 
should be used as benchmarks. Considering the 3D isotropic model ($R=1$), 
we have the inequality\cite{OZ04} $T_N>T_C$. Moreover, 
$\tilde{T}_M \equiv T_M/[|J_{\parallel}| S(S+1)]$ is found to increase with 
increasing values of $S$.\cite{OZ04,YTH05} Considering, for example, the 
RPA, those results are not reproduced, instead we have 
$\tilde{T}_N^{\text{RPA}}= \tilde{T}_C^{\text{RPA}}$, where 
$\tilde{T}_M^{\text{RPA}}$ is independent of $S$.\cite{DW94} For layered 
magnets with $R<1$, QMC and SE data in the FM case are still missing, so 
that there are no precise statements about the relation between $T_N$  and 
$T_C$ as function of the interlayer coupling. With respect to the agreement 
with the QMC and SE data, our approach represents an important improvement as 
compared, e.g., to the RPA, which is outlined in the following.
\begin{figure}[b]
   \centering
   \includegraphics[scale=0.65]{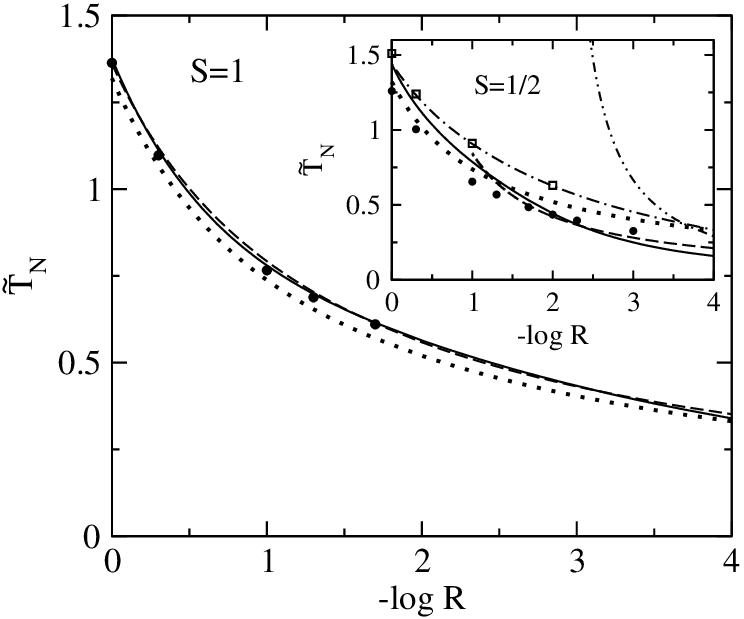}
   \caption{N\'eel temperature $\tilde{T}_N=T_N/[J_{\parallel} S(S+1)]$ 
   as a function of the interlayer coupling $R=J_{\perp}/{J_{\parallel}}$. 
   The results of the RGM (solid lines) and of the empirical formula 
   (\ref{eq_fit}) (dashed lines) are compared with the QMC data ($\bullet$, 
   Ref.~\onlinecite{YTH05}), the RPA (dotted lines, Ref.~\onlinecite{DW94}), 
   and, for $S=1/2$ (inset), with the mean-field theories of 
   Refs.~\onlinecite{IKK91} ($\square$),  \onlinecite{SI93} (dot-dashed line), 
   and \onlinecite{AA08} (dot-dot dashed line). }
   \label{fig_Tn}
\end{figure}    

For the 3D ferro- and antiferromagnets, the solution of the RGM 
self-consistency equations yields the magnetization $m(T)$ with $m(T_M)=0$ 
at the second-order phase transition temperature $T_M$, where 
$\lim_{J_{\perp} \to 0} T_M=0$ is in agreement with the 
Mermin-Wagner theorem.\cite{MW66} 
In Fig.~\ref{fig_Tn} and Table \ref{tab_2} our results for $\tilde{T}_M$  
as functions of $R$ and $S$ are presented, where in Fig.~\ref{fig_Tn} 
the N\'eel temperature $\tilde{T}_N$ is compared with the QMC data of 
Ref.~\onlinecite{YTH05} and other approaches. For the 
$S=1$ antiferromagnet we get a very good agreement with the QMC results, 
as was also found for the 2D model (see Fig.~\ref{fig_2d_afm}). Remarkably, 
the RPA results for both the $S=1$ (Ref.~\onlinecite{DW94}) and $S=1/2$ 
models\cite{DW94,MSS92} are in a rather good agreement with the QMC data. 
Considering the case $S=1/2$ (inset of Fig.~\ref{fig_Tn}) and $R<0.04$, 
we ascribe the reduction of $T_N$ found by the RGM as compared to the RPA 
and the mean-field approaches of Refs.~\onlinecite{IKK91} and 
\onlinecite{SI93} to an improved description of strong AF quantum fluctuations 
at low temperatures counteracting the formation of LRO. For further 
comparison, the N\'eel temperature given very recently\cite{AA08} by the
interlayer mean-field approach within the Schwinger-boson mean-field 
theory is depicted for $S=1/2$. The marked difference to the other curves 
(also found for $S=1$) might be due to the asymmetry between intralayer 
and interlayer correlations introduced in this approach. 
% %

Next we consider the transition temperatures $\tilde{T}_M$ for arbitrary values 
of $S$. The RGM yields $\tilde{T}_C(S) \neq \tilde{T}_N(S)$, as can be seen in 
Table \ref{tab_2}, which is in accord with the QMC and SE data, but in 
contrast to the RPA result (see above). 
In passing to the classical limit $S \to \infty$ 
we find $\lim_{S \to \infty} \tilde{T}_M = \tilde{T}_M^{RPA}$ for all 
values of $R$. This may be understood as follows. The RGM is a second-order 
theory that goes one step beyond the RPA and, therefore, provides 
a better description of quantum fluctuations. Their vanishing for 
$S \to \infty$ may be reflected in the equality of the transition temperatures.
% %
%
\begin{table*}
\caption{Coefficients of the empirical law [Eq.~(\ref{eq_fit})] for the 
transition temperatures of the ferro- and antiferromagnet.}
\begin{center}
\begin{tabular}{c|ccc|ccc|c} \hline \hline
& \multicolumn{3}{c|}{Ferromagnet} & \multicolumn{3}{c|}{Antiferromagnet} \\
\phantom{n}\phantom{n}
& \phantom{n} $S=1/2$ \phantom{n} & \phantom{n} $S=1$ \phantom{n} 
& \phantom{n} $S=3/2$ \phantom{n} & \phantom{n} $S=1/2$ \phantom{n}
& \phantom{n} $S=1$ \phantom{n} & \phantom{n} $S=3/2$ \phantom{n} 
& \phantom{n} $S=\infty$ \phantom{n} \\
\hline
$A$ & 3.15 & 4.00 & 4.27 & 1.95 & 4.36 & 4.34 & 3.96 \\
$B$ & 2.50 & 3.08 & 3.27 & 0.01 & 3.21 & 3.27 & 3.01 \\
\hline \hline
\end{tabular}
\end{center}
\label{tab_3}
\end{table*}

We compare our results for the 3D isotropic model 
($R=1$) with the SE\cite{OZ04} and QMC data \cite{YTH05} 
for different spins. For the ferromagnet, the Curie temperatures $\tilde{T}_C$ 
deviate from the SE values,\cite{OZ04}  $\tilde{T}_C=1.119$ (1.2994, 1.37) 
for $S=\frac{1}{2}$, (1, $\frac{3}{2}$), by 10\% (0.5\%, 4\%). For the 
antiferromagnet, the deviations of the N\'eel temperatures $\tilde{T}_N$ 
from the SE values\cite{OZ04} [agreeing with the QMC values for 
$S=1/2$ and $S=1$ (Ref.~\onlinecite{YTH05})], 
$\tilde{T}_N=1.259$ (1.3676, 1.404)  
for $S=\frac{1}{2}$, (1, $\frac{3}{2}$), amount to 14\%, (0.6\%, 4\%). 
From the experimental point of view, for the fit of exchange coupling 
parameters, deviations in the magnitude of transition temperatures of up to 
about 10\% are considered as a reasonable accuracy.
In both the FM and AF cases the RGM yields the best values of $\tilde{T}_M$ 
for $S=1$. For any spin, we get the correct relation $T_N>T_C$, where the 
ratio $Q=T_N/T_C=1.17$ (1.05, 1.02) for $S=\frac{1}{2}$, (1, $\frac{3}{2}$) 
agrees well with the SE values $Q=1.13$ (1.05, 1.03). That means, concerning 
the difference between $T_N$ and $T_C$, the RGM yields good results for all 
values of $S$. Considering the dependences $\tilde{T}_M(S)$, the increase of  
 $\tilde{T}_C$ with increasing $S$ is in qualitative agreement with the 
 SE data. For the antiferromagnet, $\tilde{T}_N$ decreases with increasing 
 $S$ being opposite to the behavior of the SE\cite{OZ04} and QMC 
 data.\cite{YTH05} This is connected with the inequality 
 $\tilde{T}_N (S=\frac{1}{2}) > \lim_{S \to \infty} \tilde{T}_N=1.3189$, 
 whereas the QMC data\cite{YTH05, HJ93} yield 
$\tilde{T}_N (S=\frac{1}{2}) < \tilde{T}_N (S=\infty)=1.443$ [note that in the 
classical Heisenberg model\cite{HJ93} the spins are taken of unit length, and 
the exchange interaction $J^{cl}$ is related to 
$ J \equiv J_{\perp} = J_{\parallel}$ by $J^{cl}=JS(S+1)$].

Let us consider the anisotropic magnets ($R<1$). For $S=1/2$ and $R<0.01$ 
we find $T_N <T_C$, and for $R \geqslant 0.01$ we have $T_N > T_C$. 
In the cases $S=1$ and $S=3/2$ we get $T_N>T_C$ 
for all values of $R$. 
The peculiarity in the relation between $T_N$ and $T_C$ for $S=1/2$ may be 
explained by the presence of strong AF quantum fluctuations at low 
temperatures which may suppress the AF LRO.

For the discussion of experimental data it is convenient to use an 
analytical expression for $T_M(R,S)$. 
Our RGM results for the dependence of $\tilde{T}_{M}$ on $R$ 
may be well fitted by the empirical formula proposed in 
Ref.~\onlinecite{YTH05}, 
\begin{equation}
\tilde{T}_M=\frac{A}{B-\ln(J_{\perp}/J_\parallel)},
\label{eq_fit}    
\end{equation}
where the values of $A$ and $B$ are listed in Table \ref{tab_3}. The 
concrete values of the coefficients slightly depend on the choice of data 
points used for the fit. 
Since $T_M$ reveals the strongest increase with $R$ for $R \ll 1$, 
in this region we take points lying more dense than for moderate 
interlayer couplings. 
The values given in Table \ref{tab_3} are 
obtained by choosing points within the interval $R=10^{-4}$ to 
$10^{-2}$ and $R=10^{-2}$ to 1 in steps of $\Delta R = 10^{-4}$ 
and $\Delta R=10^{-2}$, respectively. 
Then, a good fit in the whole $R$ region can be achieved in all cases, 
except for the $S=1/2$ antiferromagnet, where a reasonable fit by 
Eq.~(\ref{eq_fit}) is obtained for $R \leqslant 0.1$ (see Fig.~\ref{fig_Tn}).
% %
% %
\subsection{Specific heat}\label{subsec_fajho}
\begin{figure}[b]
   \centering
   \includegraphics*[scale=0.65]{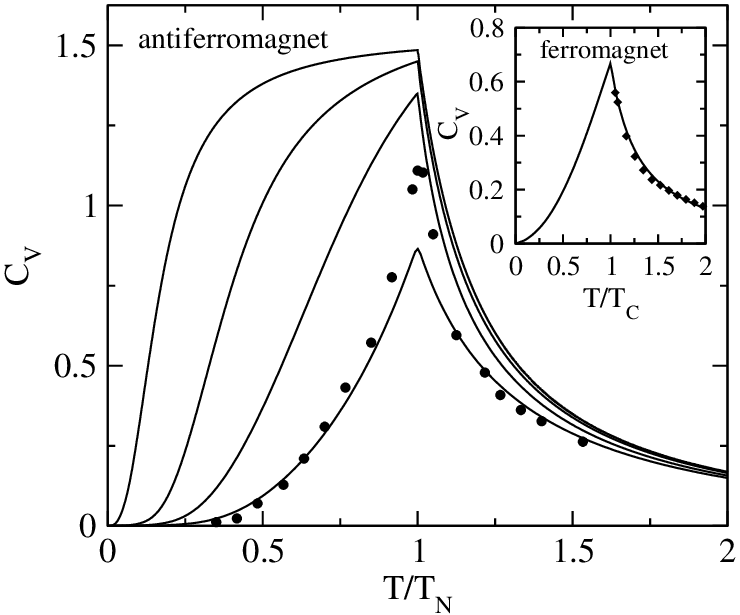}
   \caption{Specific heat $C_V$ of the isotropic antiferromagnet 
   ($J_{\perp}=J_{\parallel}$) with $S=1/2$, 1, 2, and 5, from bottom 
   to top, in comparison to the QMC data for $S=1/2$ 
   ($\bullet$, Ref.~\onlinecite{San98}). The inset displays $C_V$ for the 
   isotropic $S=1/2$ ferromagnet, compared to the SE results of 
   Ref.~\onlinecite{OB96} ($\blacklozenge$).}
   \label{fig_4}
\end{figure}    
\begin{figure}[b]
   \centering
   \includegraphics*[scale=0.65]{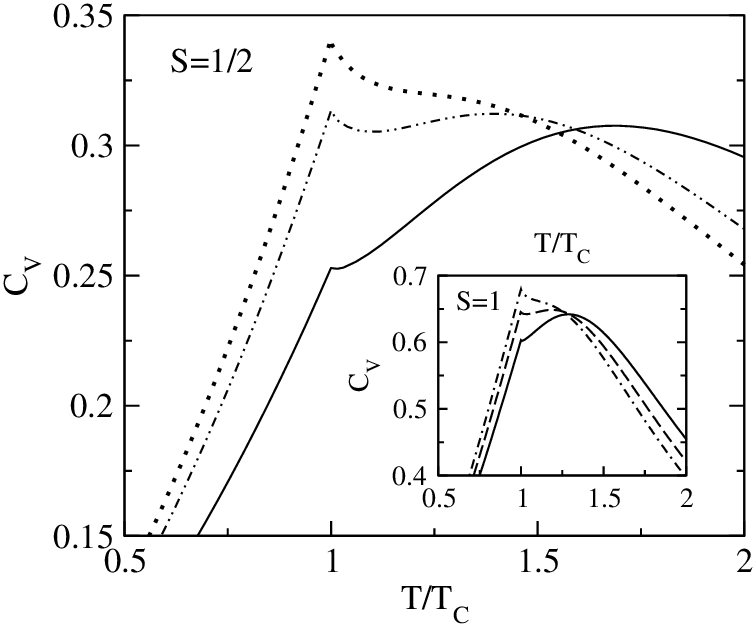}
   \caption{Specific heat $C_V$ of the ferromagnet with $S=1/2$ and $S=1$ 
   (inset) for $J_{\perp}/J_{\parallel}=0.01$ (solid lines), with 
   $S=1/2$ for $J_{\perp}/J_{\parallel}=0.025$ (dot-dot dashed line) and 
   0.035 (dotted line), and with $S=1$ for  $J_{\perp}/J_{\parallel}=0.015$ 
   (dashed line) and 0.02 (dot-dashed line).}
   \label{fig_5}
\end{figure}    
The temperature dependence of the specific heat $C_V$ is characterized by a 
cusplike singularity at the transition temperature $T_M$ 
determined by $J_{\perp}$ and, for sufficiently 
low interlayer couplings, by a broad maximum above $T_M$ 
that is mainly determined by $J_{\parallel}$. 
For the 3D isotropic 
magnets, $C_V$ is plotted in Fig.~\ref{fig_4}. Considering the $S=1/2$ 
ferromagnet (see inset), above $T_C$  we obtain an excellent agreement with 
the SE data of Ref.~\onlinecite{OB96}. For the $S=1/2$ antiferromagnet, 
the agreement 
of the RGM with the QMC results\cite{San98} is very good at temperatures 
sufficiently below and above $T_N$, whereas near $T_N$ the height 
of the cusp is underestimated. 
Considering the $S$ dependence of $C_V$ in the LRO phase, with 
increasing $S$ the 
slope of the $C_V$ curves near $T_N$ decreases, and the cusp develops 
to a kink 
(see Fig.~\ref{fig_4}). The analogous tendency is found in the FM case. 
This behavior may be considered as a deficiency of the RGM, because in the 
classical Heisenberg model ($S \to \infty$) the QMC data of 
Ref.~\onlinecite{HJ94} yield evidence for a cusplike structure of $C_V$ at 
$T_M$.

\begin{figure}
   \centering
   \includegraphics*[scale=0.65]{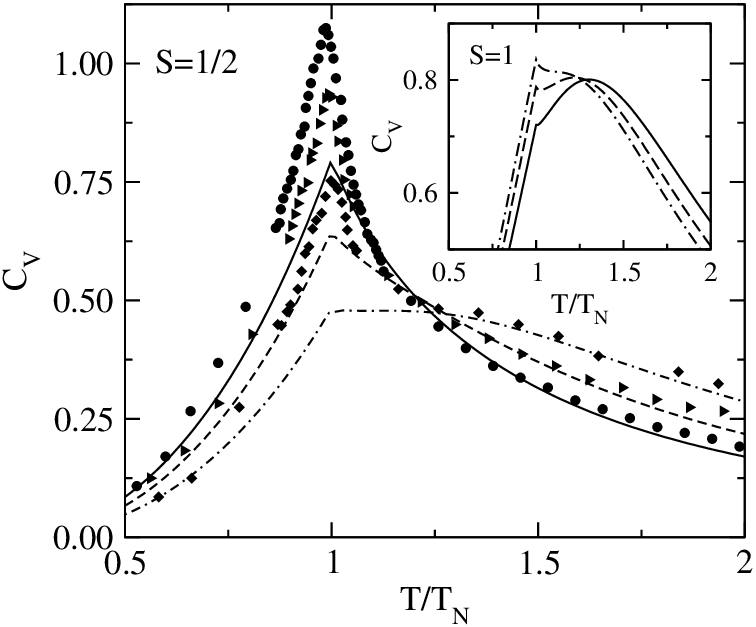}
   \caption{Specific heat of the $S=1/2$ antiferromagnet in comparison to 
   the QMC data of Ref.~\onlinecite{SSS03} (filled symbols) for 
   $J_{\perp}/J_{\parallel}=2^{-1}$ (solid line, $\bullet$), $2^{-2}$ 
   (dashed line, $\blacktriangleright$), 
   and $2^{-3}$ (dot-dashed line, $\blacklozenge$). 
   The inset shows $C_V$ of the $S=1$ antiferromagnet for 
   $J_{\perp}/J_{\parallel}=0.01$ (solid line), 0.015 (dashed line), and 
   0.02 (dot-dashed line).}
   \label{fig_6}
\end{figure}

Next we consider the specific heat of quasi-2D magnets. In the ferromagnet 
a broad maximum, in addition to the phase-transition singularity, appears at 
$R<0.035$ ($R \leqslant 0.015$) for $S=1/2$ ($S=1$), 
as can be seen in Fig.~\ref{fig_5}. The 
analogous behavior is found for the antiferromagnet, as shown in 
Fig.~\ref{fig_6}. Here, the broad maximum occurs at $R<2^{-3}$ 
($R \leqslant 0.015$) for $S=1/2$ (S=1), which agrees with the $S=1/2$ QMC 
data of Ref.~\onlinecite{SSS03}. As for the isotropic $S=1/2$ antiferromagnet 
(cf.~Fig.~\ref{fig_4}), the RGM agrees well with the QMC results at low 
and high temperatures. 
Again, the height of the cusp is underestimated, 
where the relative deviation of $C_V(T_N)$ from the QMC values increases with 
decreasing $R$.
% %
% %
%

Recently, specific heat data for the quasi-2D $S=1/2$ antiferromagnet 
Zn$_2$VO(PO$_4$)$_2$ were presented.\cite{KKG06} 
Taking $T_N=3.75$K and $J_{\parallel}=7.41$K from Ref.~\onlinecite{KKG06},
by Eq.~(\ref{eq_fit}) and Table \ref{tab_3} we get 
$R=5.8 \times 10^{-2}$. Calculating the specific heat 
we obtain a broad maximum at $T_m=5.9$K with $C_V(T_m)=0.45$ which 
corresponds to the measured broad hump at $T_h=4.5$K with the height 
$C_V(T_h)=0.45$ agreeing with the theoretical value of $C_V(T_m)$.
At $T_N$, the experiment shows a pronounced cusp with $C_V(T_N) \simeq 0.6$. 
As discussed above (see Fig.~\ref{fig_6}), this feature cannot be reproduced 
by the RGM, instead we get a small spike at $T_N$ with $C_V(T_N) \simeq 0.3$. 
\subsection{Correlation length}\label{subsec_korrlength}
\begin{figure}%[b]
   \centering
    \includegraphics*[scale=0.65,clip=true]{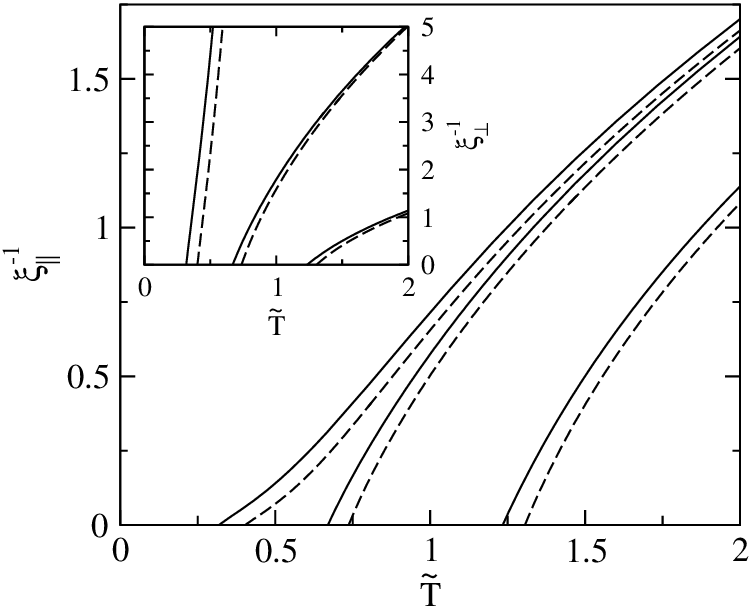}
   \caption{Inverse correlation lengths within ($\xi_{\parallel}^{-1}$) 
   and between the xy planes ($\xi_{\perp}^{-1}$, see inset) 
   versus $\tilde{T}=T/[|J_{\parallel}|S(S+1)]$ of the 
   ferromagnet with $S=1/2$ (solid lines) and 
   $S=1$ (dashed lines) for $J_{\perp}/J_{\parallel}=0.001$, 0.1, and 1, 
   from left to right.}
   \label{fig_7}
\end{figure}
\begin{figure}[b]
   \centering
   \includegraphics[scale=0.65]{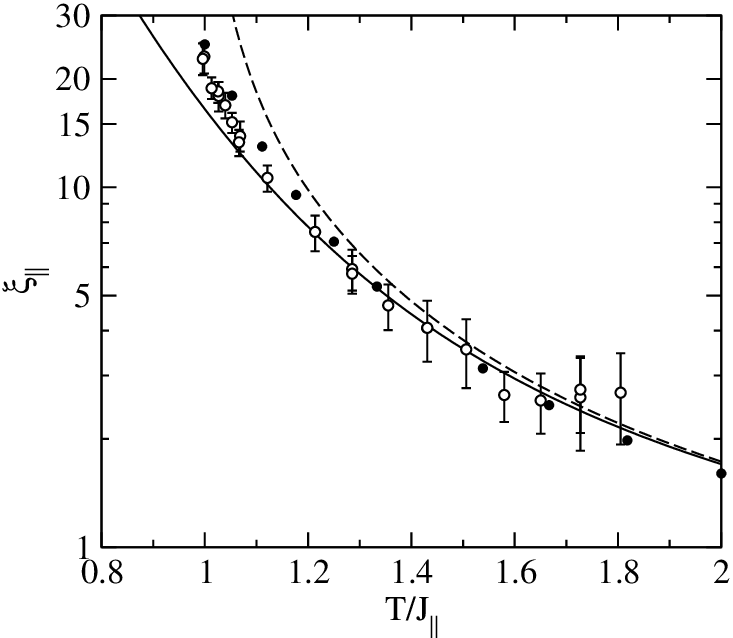}
   \caption{Antiferromagnetic intralayer correlation length in 
   La$_2$NiO$_4$ obtained by the neutron-scattering experiments of 
   Ref.~\onlinecite{NYH95} ($\circ$) compared to the QMC data for 
   $J_{\perp}=0$ ($\bullet$, Ref.~\onlinecite{HTK98}) and the RGM 
   results for  $J_{\perp}=0$ (solid line) and 
   $J_{\perp}/J_{\parallel}=3.5 \times 10^{-3}$ (dashed line).                                                                                                                                                             
   }
   \label{fig_8}
\end{figure}
The intralayer and interlayer correlation lengths $\xi_{\mu}$, 
($\mu=\parallel, \perp$) for  $R \neq 0$ 
 diverge as $T$ approaches $T_M$ from above. In 
the vicinity of $T_M$, $\xi_{\parallel}^{-1}$ and $\xi_{\perp}^{-1}$ 
behave as $T-T_M$ (corresponding to the critical index $\nu=1$) also found 
by previous mean-field approaches.\cite{SI93,IKK99} 
This can be seen in Fig.~\ref{fig_7} that shows $\xi_{\mu}^{-1}$ versus 
$\tilde{T}=T/[|J_{\parallel}|S(S+1)]$ of the $S=1/2$ and $S=1$ ferromagnet. 
The curves for the antiferromagnet look similar. 
At fixed $R<1$ and 
$S$ we have $\xi_{\perp} < \xi_{\parallel}$ which corresponds to the weaker 
interlayer as compared to the intralayer correlations. 
Considering the $S$ dependence of $\xi_{\mu}$ for the ferromagnet, we 
have $\tilde{T}_C(S=1/2)<\tilde{T}_C(S=1)$ (see Fig.~\ref{fig_7} and 
Sec.~\ref{subsec:Tcn}) which implies, at fixed $\tilde{T}>\tilde{T}_C$ 
and $R$, the 
inequality $\xi_{\mu}(S=1/2)<\xi_{\mu}(S=1)$. Note that recently, an analogous 
$S$ dependence for the lonitudinal correlation length $\xi^{zz}$ of the 
2D ferromagnet in a small magnetic field was found, also by QMC,\cite{JIB08} 
i.e., $\xi^{zz}(S=1/2)<\xi^{zz}(S=1)$ at fixed $\tilde{T}$. 

Let us compare our results for the intralayer correlation 
length $\xi_{\parallel}$ 
with the neutron-scattering data on the $S=1$ quasi-2D antiferromagnet 
La$_2$NiO$_4$.\cite{NYH95} Taking $T_N=327.5$K and $J_{\parallel}=28.7$meV 
from Ref.~\onlinecite{NYH95}, by Eq.~(\ref{eq_fit}) and Table \ref{tab_3} 
we obtain $R=3.5 \times 10^{-3}$. In Fig.~\ref{fig_8} 
the experimental data are plotted in comparison to the QMC data for 
$R=0$ (Ref.~\onlinecite{HTK98}) and the RGM results 
for $R=0$ and 
 $R=3.5 \times 10^{-3}$, where a satisfactory overall 
 agreement with experiments is found. At fixed temperature, the correlation 
length for $R>0$ is larger than for $R=0$, because 
$\xi_{\parallel}$ diverges at $T_N$. To explain the 
neutron-scattering experiments, 
in Ref.~\onlinecite{NYH95} a small Ising anisotropy in the strictly 2D 
model was considered which leads to in a finite transition 
temperature 
somewhat below $T_N$. Such an easy-axis anisotropy was also discussed 
in Ref.~\onlinecite{GP01} to explain the experiments. However, as was shown 
in Ref.~\onlinecite{HTK98}, the experimental data with 
$\xi^{\text{exp}}<\xi^{\text{QMC}}$ (see Fig.~\ref{fig_8}) are incompatible 
with the QMC results 
obtained for the 2D model with 
a small Ising anisotropy, since it even  
enhances the correlation length at low temperature. In our approach, the 
finite value of $T_N$ is ascribed entirely to the interlayer coupling which 
gives $\xi > \xi^{\text{QMC}}$. To improve the agreement with experiments, 
let us point out, that in our calculations a simple cubic lattice was taken, 
whereas in the orthorhombic structure of La$_2$NiO$_4$ the interlayer coupling 
is frustrated. As was shown in Ref.~\onlinecite{HRI08}, in the 
$J_1-J_2$ model, frustration may appreciably reduce the correlation length. 
The influence of frustration on the transition temperature and 
correlation length of quasi-2D Heisenberg magnets will be left for further 
study.  
 \section{SUMMARY}\label{sec:sum}
In this paper the thermodynamics of layered Heisenberg magnets with arbitrary 
spin $S$ is systematically investigated by a spin-rotation-invariant 
Green-function method and by exact diagonalizations on finite 2D lattices. 
The main focus is put on the calculation of the Curie temperature $T_C$ 
and the N\'eel temperature $T_N$ in dependence on the interlayer coupling 
$J_{\perp}$ and the spin quantum number. From the numerical data we obtain simple 
empirical formulas for $T_{C,N} (J_{\perp})$.  
A good agreement of our results, in particular on the relation 
between $T_C$ and $T_N$, with available quantum Monte Carlo and 
series-expansion data is found. The comparison to experiments on the 
quasi-2D antiferromagnets Zn$_2$VO(PO$_4$)$_2$ and La$_2$NiO$_4$ yields a 
reasonable agreement. From our results we conclude that the application 
of the second-order Green-function approach to extended layered Heisenberg 
models (frustration, anisotropy in spin space) may be promising to describe the 
unconventional magnetic properties of real low-dimensional quantum spin systems.

\end{document}